\documentstyle[12pt,amssymb,amsfonts,amsbsy]{article}
\newtheorem{definition}{Definition}
\newtheorem{theorem}{Theorem}
\begin{document}
\author{Ilja Schmelzer\thanks
       {WIAS Berlin}}

 \title{Generalization Of Lorentz-Poincare Ether Theory To Quantum
Gra\-vity}

\begin{abstract}
\sloppypar

We present a quantum theory of gra\-vity which is in agreement with
observation in the relativistic domain.  The theory is not
relativistic, but a Galilean invariant generalization of
Lorentz-Poincare ether theory to quantum gra\-vity.

If we apply the methodological rule that the best available theory has
to be preferred, we have to reject the relativistic paradigm and
return to Galilean invariant ether theory.

\end{abstract}

\maketitle

\begin{center}{\it
... die blo{\ss}e Berufung auf k\"{u}nftig zu entdeckende Ableitungen
bedeutet uns nichts.} \hfill Karl Popper
\end{center}

\begin{center}{\it In quantum gra\-vity, as we shall see, the
space-time manifold ceases to exist as an objective physical reality;
geometry becomes relational and contextual; and the foundational
conceptual categories of prior science -- among them, existence itself
-- become problematized and relativized.} \hfill Alan Sokal
\end{center}

\tableofcontents

\section{The Problem Of Quantum Gra\-vity}

We believe that there exists a unique physical theory which allows to
describe the entire universe.  That means, there exists a theory ---
quantum gra\-vity --- which allows to describe quantum effects as well
as relativistic effects of strong gravitational fields.

The simplest strategy would be to search for a theory which is in
agreement with the metaphysical principles in above limits --- a
relativistic quantum theory of gra\-vity.  This theory

 \begin{itemize}
 \item allows to derive the experimental predictions of general
relativity in $\hbar \to 0$ and Schrvdinger theory in $c \to \infty$;
 \item is a quantum theory;
 \item is in agreement with the relativistic paradigm.
 \end{itemize}

The strategy seems to fail.  After a lot of research we have, instead
of a theory, a list of serious problems: problem of time, topological
foam, non-renormalizability, information loss problem.  The
consequences of relativistic quantum gra\-vity seem close to Sokal's
parody \cite{Sokal}.

What if this strategy really fails?  In this case, at least one of the
two paradigms of modern science has to be wrong.  This makes quantum
gra\-vity --- a theory we possibly never need to predict real
experimental results --- very interesting for current science.

But, if one of the principles is wrong, how can we find out which?  To
find the answer, we can apply standard scientific methodology
(following Popper \cite{Popper}).  All we need is the following
methodological rule: We always have to prefer the best available
theory, to refer to possible results of future research is not
allowed.  What we have to do is to present one of the following
theories:

 \begin{itemize}

 \item post-relativistic quantum theories (non-relativistic theories
which predict relativistic effects in some limit);

 \item post-quantum relativistic theories (non-quantum theories which
but predict quantum effects in some limit).

 \end{itemize}

Until relativistic quantum gra\-vity has not been found, the principle
which is not valid in this theory is the wrong one.  Indeed, in this
case we have a theory which fulfils our first condition.  We have to
prefer this theory as the best available theory, and reference to the
future success of relativistic quantum gra\-vity is not allowed. qed.

The rejected paradigm may be revived by the results of future
research.  But this is a trivial remark --- it is correct for every
invalid paradigm.  Thus, this paradigm is as dead as possible for a
scientific paradigm.

This result depends on some methodological rules we have to accept.
We have to make a decision to apply a certain methodology of empirical
science.  We need:

 \begin{itemize}

 \item The rule that we have to prefer the best available theory and
to ignore the hope for future success of alternative approaches;

 \item A rule which allow to prefer a unified theory of quantum
gra\-vity compared with theories which do not allow to describe
quantum effects of strong gravitational fields;

 \end{itemize}

Our decision was to accept the methodology of Popper \cite{Popper}.
It contains the rule that we have to prefer the best available theory.
Popper's criterion of potential predictive power obviously prefers a
unified theory of quantum gra\-vity.

\section{Introduction}

The aim of this paper is to present a post-relativistic quantum theory
of gra\-vity --- that means, a non-relativistic theory which
nonetheless describes relativistic effects correctly.

Our strategy may be described as the simplest search strategy after
the search for relativistic quantum gra\-vity.  We have to omit at
least one of the guiding principles --- the relativity principle or
quantum theory.  As guiding principles, we use the other principle and
an available competitor of the rejected principle.

For quantum theory we have no appropriate known competitor --- Bohm's
hidden variable theory seems to be even less compatible with
relativity.  For the relativity principle the best known
non-relativistic theory is Lorentz-Poincare ether theory. It is
Galilean invariant, thus, we have obviously no ``problem of time''. We
have a fixed, flat space, thus, no topological foam.  Thus, to replace
relativity by the Lorentz-Poincare ether paradigm solves at least two
of the quantization problems.

Thus, we try to use the pre-relativistic ether paradigm as the
replacement of the relativity principle and search for a theory of
gra\-vity which

 \begin{itemize}

 \item is a quantum theory,

 \item is Galilean invariant,

 \item allows to describe relativistic time dilation as caused by
interaction with a physical field --- the ``ether''.

 \end{itemize}

We already know how we have to describe special-relativistic effects,
thus, what we have to do is to generalize this scheme to gra\-vity and
to quantize the resulting theory in the simplest possible way --- with
canonical quantization.

To realize this concept is surprisingly simple.  

We know that general relativity is tested for a wide range, thus, the
simplest idea is to remain as close as possible to general relativity.
Thus, it would be the simplest case if we have a Lorentz metric
$g_ij(x,t)$ in our theory and this metric fulfills the Einstein
equations.  What we have to include is a preferred Newtonian frame.
The simplest idea is to choose harmonic coordinates to define this
frame.

We obtain a theory --- post-relativistic gra\-vity --- which is
slightly different from general relativity, with a natural ether
interpretation: we can identify a ``density'' $\rho$, a ``velocity''
$v^i$, and a ``stress tensor'' $\sigma^{ij}$ with correct
transformational behaviour and the usual conservation laws.  The
simplest way to incorporate the harmonic equation into the Lagrange
formalism requires to break the covariance.  This makes the ether
observable in principle --- as some type of dark matter, with two
cosmological constants which have to be fitted by observation.

Once we have an ether theory, the question if this ether has an atomic
structure is natural.  Is our continuous ether theory valid for
arbitrary distances or is it only the large scale approximation of
some atomic ether theory for distances below some $\epsilon$?  The
first hypothesis leads to problems already in classical theory, but in
the quantum case we obtain ultraviolet problems.

That's why we assume that the atomic hypothesis is true.  This
obviously solves the ultraviolet problems.  We show this for a simple
example theory with discrete structure --- standard regular lattice
theory.  For this theory, standard canonical quantization may be
applied without problem.  This suggests that for better atomic
theories canonical quantization works too.

Our lattice theory is not the ideal atomic ether theory.  But this is
already a Galilean invariant quantum theory of gra\-vity, and it works
in the relativistic domain.  Thus, it already solves the problem of
quantum gra\-vity.  In other words, the solution of the ``problem of
quantum gra\-vity'' can be described in a single sentence:

\begin{center}{\it Classical canonical Weyl quantization of a lattice
theory (standard first order finite elements on a regular rectangular
lattice in space) of a classical Galilean invariant theory for an
``ether'' $g_{ij}$ and matter fields $\phi$ with Lagrangian

\[ L 	= L_{Rosen}(g_{ij})
	+ L_{matter}(g_{ij},\phi)
	+ \lambda_1 g^{00}
	+ \lambda_2 g^{aa}
\]

}\end{center}

Note that to remove the regularization, that means to solve the
ultraviolet problem by some sort of renormalization, is necessary only
in a relativistic theory.  The point is that the regularization is not
relativistic invariant.  That's why this theory is far away from being
``relativistic quantum gra\-vity''.

But Galilean invariant regularization is not problematic, and that's
why this simple theory already solves the problem of Galilean
invariant, non-relativistic quantization of gra\-vity, even for the
relativistic domain.

\section{Generalization Of Lorentz-Poincare Ether Theory To Gra\-vity}

In this section, we present the details of the definition of our ether
theory.  We present a slightly more general scheme, which shows that
our approach is not very much related with the details of the theory
of gra\-vity.

\begin{definition} 
Assume, we have a classical relativistic theory with the following
variables: the Lorentz metric $g_{ij}$ and some matter fields
$\phi^m$, with a relativistic Lagrangian

\[ L_{rel} = L_{rel}(g_{ij},g_{ij,k},\phi^m,\phi^m_{,k}) \]

In this case, the ``related ether theory'' is defined in the following
way:

The theory is defined on Newtonian space-time $R^3 \times R$. The
independent variables are:

 \begin{itemize}

 \item a positive scalar field $\rho(x,t)$ named ``density of the
ether'',

 \item a vector field $v^a(x,t), 1 \le a \le 3$ named ``velocity of
the ether'',

 \item a positive-definite symmetrical tensor field $\sigma^{ab}(x,t),
1 \le a,b \le 3$ named ``stress tensor of the ether'',

 \item and one field $\phi^m(x,t)$ for every ``matter field'' of the
original relativistic theory named ``inner step of freedom of the
ether''.

 \end{itemize}

The Lagrange functional is

\[ L 	= L_{rel}(g_{ij},g_{ij,k},\phi^m,\phi^m_{,k})
	+ \lambda_1 g^{00} + \lambda_2 g^{aa}
\]

where the ``ether metric'' $g_{ij}(x,t)$ is defined by the following
formulas:

\begin{eqnarray*} \label{gdef}
 \hat{g}^{00} &= g^{00} \sqrt{-g} =  &\rho \\
 \hat{g}^{a0} &= g^{a0} \sqrt{-g} =  &\rho v^a \\
 \hat{g}^{ab} &= g^{ab} \sqrt{-g} =  &\rho v^a v^b - \sigma^{ab} 
\end{eqnarray*}

\end{definition}

The most interesting example is of course general relativity.  The
related ether theory we have named post-relativistic gra\-vity or
simply post-relativity.  In the following we restrict ourself to this
theory.  Nonetheless, the previous scheme may be applied to other
metrical theories of gra\-vity too.  That means, the question of
existence of an ether does not depend on the details of the theory of
gra\-vity.

\subsection{Elementary Properties Of Post-Relativistic Gra\-vity}

The ``ether metric'' in post-relativistic gra\-vity identifies the
ether with the gravitational field.  It is in general inhomogeneous
and instationary.  Only in the Minkowski limit the ether becomes
homogeneous and stationary, and obviously coincides with the ether of
Lorentz-Poincare ether theory.  Because of our assumptions about
$\rho$ and $\sigma^{ab}$, $g_{ij}$ is really of signature $(1,3)$.

The ether influences all matter fields only via minimal interaction
with the ether metric, thus, in the same way.  Thus, we have no ether
for the electro-magnetic field only, but a common ether for all matter
fields.  Especially, all clocks --- which have to be described by
matter fields --- are time-dilated in the same way and show
general-relativistic proper time $\tau$.  But this ``proper time''
does not have the metaphysical status of time.  It is only a parameter
for the speed of clocks, which depends on the state of the ether and
the relative velocity of the clock and the ether.

Our Lagrange density is not covariant. To compare it with relativistic
theory, it is useful to introduce the preferred coordinates as
independent functions $T, X^a$.  This allows to make the Lagrange
density covariant:

\[ L 	= L_{rel}(g_{ij},g_{ij,k},\phi^m,\phi^m_{,k})
	+ \lambda_1 g^{ij}T_{,i}T_{,j} 
	+ \lambda_2 g^{ij}X^a_{,i}X^a_{,j}
\]

Now we see that the relativistic field equations are fulfilled, with a
small modification --- an additional term has to be added to the
energy-momentum tensor.  In the preferred coordinates $T, X^a$, the
additional energy term is

\[(T_{full})^0_0 = (T_{rel})^0_0 + \lambda_1 g^{00} - \lambda_2
g^{aa}.\]

For the absolute coordinates, we obtain the usual harmonic wave
equation

\[ \Box T = 0;\;\; \Box X^a = 0 \]

Translated into the original variables, these are simply conservation
laws for the ether:

\[ \partial_t \rho + \partial_i (\rho v^i) = 0 \]

\[ \partial_t (\rho v^j) + \partial_i(\rho v^i v^j - \sigma^{ij}) = 0
\]

\subsection{The Cosmological Constants}

Thus, the only immediately observable difference between general
relativity and post-relativistic gra\-vity are two unknown constants
$\lambda_1, \lambda_2$.  To observe them we have to compare the
energy-momentum tensor of observable matter with the observable
Einstein tensor in harmonic coordinates.  The observed difference
should have the form $\lambda_1 g^{00} -\lambda_2 g^{aa}$.  These
parameters may be considered as cosmological constants of
post-relativistic gra\-vity.

This does not mean that this observation immediately falsifies
relativity.  We can explain the same observation inside the
relativistic paradigm too.  Indeed, we can interpret the fields $T,
X^a$ as some new matter fields.  They obviously fulfil a relativistic
equation.  The fields $T, X^a$ are defined only modulo a constant,
thus, describe only some potential, only the derivatives
$\hat{g}^{ij}T_{,j},\hat{g}^{ij}X^a_{,j}$ are physical fields.  These
fields do not interact with other matter.  Thus, all what we observe
from relativistic point of view is more or less standard dark matter.

\subsubsection{The Observation Of The Cosmological Constants}

To observe the cosmological constant seems possible for the global
homogeneous universe solution.  In post-relativistic gra\-vity, only
the flat universe is homogeneous --- a preference of the flat universe
which is supported by observation.  Now, let's look how dark matter of
the form $\lambda_1 g^{00} -\lambda_2 g^{aa}$ in harmonic coordinates
modifies the flat Friedman solution (c=1)

\[ ds^2 = d\tau^2 - b^2(\tau)(dx^2+dy^2+dz^2). \]

The harmonic coordinates are $x,y,z$ and $t=\int b^{-3}(\tau)d\tau$,
thus, we have 

\[ ds^2 = b(t)^6 dt^2 - b^2(t)(dx^2+dy^2+dz^2). \]

The first part $\lambda_1 g^{00}\sqrt{-g} = \lambda_1$ is a fixed
density --- something like invisible dust in rest.

The second part $-\lambda_2 g^{aa}\sqrt{-g} = -3\lambda_2b^4(t)$ has a
different behaviour.  The parameter $\lambda_2$ obviously influences
the observable (proper time) age of the universe.

In other words, if the cosmological constants are non-trivial and have
observable effects, we will observe:

 \begin{itemize}

 \item a missing mass --- a difference between observed mass and mass
necessary to explain the current Hubble coefficient in general
relativity.

 \item a wrong age of the universe --- a difference between the
observable age of the universe and the age which follows from general
relativity and the current Hubble coefficient.

 \end{itemize}

The observations may be described in general relativity as more or
less standard dark matter.

\subsubsection{The Necessity Of Cosmological Constants}

We know that there are problems with missing mass and the age of the
universe.  We have to admit that this knowledge has influenced our
decision to incorporate these cosmological constants into
post-relativistic gra\-vity.  But, remember the history of the
cosmological constant in general relativity. Is it possible that the
introduction of these constants is an error in post-relativity too?

For an ether interpretation of $g^{ij}$ we need conservation laws ---
the harmonic condition.  To incorporate the harmonic condition into
general relativity without cosmological constants we have some
possibilities: we can use penalty terms like $g_{ij}\Gamma^i\Gamma^j$,
or Lagrange multipliers like $\lambda_i\Gamma^i$.  We can also
consider the harmonic equation as an external constraint, which does
not have to be derived with the Lagrange formalism.  Indeed,
non-harmonic configurations violate conservation laws, thus, are
simply meaningless, not part of the physical configuration space.
Variation in these directions is meaningless.

Comparing these possibilities, we nonetheless tend to prefer the
variant with cosmological constants, because of the following
arguments:

 \begin{itemize} 

 \item They seem to be the simplest way to incorporate the harmonic
equation into the theory.

 \item The related terms $\rho$ and $\rho v^2 - \sigma^{aa}$ look much
more natural from point of view of an ether interpretation, if we
compare them with the other possibilities like
$g_{ij}\Gamma^i\Gamma^j$.

 \item They break relativistic symmetry and make the ``hidden
preferred frame'' observable.  This allows to avoid classical
positivistic argumentation against hidden variables.\footnote{I have
to remark that I consider this argumentation, as well as positivism,
as invalid.  Moreover, there are two other domains where the hidden
variables become observable: for small distances it follows from the
atomic hypothesis, and in the quantum domain space-time points have an
invariant meaning, which also violates relativistic symmetry
requirements.}

 \end{itemize}

Thus, there are some independent reasons to prefer the theory with
cosmological constants.  

In the following, we ignore the questions related with the
cosmological constants.  That means, we consider the theory in the
domain with $\lambda_1,\lambda_2\approx 0$, in other words, we
consider solutions of general relativity in harmonic coordinates as
solutions of post-relativistic gra\-vity.  In this domain, we have
relativistic symmetry for all observables, the ``preferred frame'' is
a hidden variable.

\subsection{The Global Universe}

Let's consider now what happens with the global universe.  For this
purpose, we have to consider homogeneous solutions of
post-relativistic gra\-vity.

In the case without cosmological constants, we have to use homogeneous
solutions of general relativity and to introduce homogeneous harmonic
coordinates.  This is possible only for the flat universe.  Thus, that
the universe is flat is a consequence of the assumption that it is
homogeneous in the large scale.

For the standard Friedman solution

\[ds^2= d\tau^2 - \tau^{4/3}(dx^2+dy^2+dz^2) \]

we obtain the absolute (harmonic) coordinates $x,y,z$ and
$t=-\tau^{-1}$.  This leads to the following interpretation:

 \begin{itemize}

 \item All galaxies remain on it's true place, the ether density
remains constant.

 \item The observed expansion is explained by distortion of rulers. 
All our rulers become smaller.

 \item The limit $\tau\to\infty$ may be reached in finite absolute
time.

 \end{itemize}

Thus, instead of the big bang singularity in the past, we have now a
singularity in future.  We don't know yet if this future singularity
is only a property of this particular solution or a general problem.
In the following, we assume the worst case --- that this singularity
is as unavoidable as the big bang and black hole singularity in
general relativity.

An interesting property of this singularity is that our rulers become
infinitely small in absolute distances and constant ether density.
This suggests a simple solution of this problem --- an atomic
hypothesis.  If the ether has some atomic structure, in some future
the rulers become small enough to observe these atoms.  At this
moment, the application of continuous theory is no longer justified,
we have to use the atomic theory to understand what happens with our
rulers.  Thus, before the singularity $\tau\to\infty$ happens,
continuous theory breaks down and has to be replaced by an atomic
theory.

\subsection{Gravitational Collapse}

For the consideration of solutions with special symmetries, like
spherical symmetry, we also have no problem of choice of the harmonic
coordinates --- we require that they have to be symmetric too.  For
example, for the solution of a spherical static star we have the
following harmonic coordinates:

\[ ds^2 = (1-{mm'\over r})
		({r-m\over r+m}dt^2-{r+m\over r-m}dr^2)
	- (r+m)^2 d\Omega^2 \]

Here $m=m(r)=GM(r)/c^2$, $M(r)$ is the mass inside the sphere of
radius r.  For a collapsing star the situation is a little bit more
complicate, nonetheless we have a unique choice --- no incoming waves.
This has been found already by Fock \cite{Fock}.  A spherical
symmetric collapse or explosion leads to outgoing ether density waves.

A remarkable property of post-relativistic gra\-vity is that the part
behind the horizon is not part of the complete solution.  In the ether
interpretation, the time dilation for a falling observer becomes
infinite, in a way that the limit of observed, distorted, time remains
finite in infinite absolute time.  Thus, in post-relativity we have no
``black hole'' in the general-relativistic sense.  The part behind the
horizon is not physical.  The old notion ``frozen star'', which is
also in agreement with the temperature of Hawking radiation, seems
more appropriate.

Note that ether density is greater near the surface, and reaches
infinity at the horizon.  The collapse leads to a flow of ether into
the black hole.  This suggests that the theory may break down already
some time before horizon formation.  Probably, once the ether has
reached some critical density, the collapse stops.

This leads to a theoretical possibility to test the theory: fall into
a great black hole like the center of the galaxy. In general
relativity, we will reach the region behind the horizon and die in the
singularity. In post-relativistic gra\-vity, we stop falling before
the horizon, and survive there up to the time of the breakdown of the
continuous approximation we have found for the global universe.

This is of course far away from realization. In the domain where we
have experimental data, post-relativistic gra\-vity coincides with
general relativity.  This includes not only tests of the PPN
parameters in the solar system, but also tests for strong
gravitational fields like black holes outside the horizon.

\subsection{A Post-Relativistic Lattice Theory}

We have observed internal problems of the continuous theory which may
be interpreted as a breakdown for small distances --- a solution which
leads to infinite values in finite time.  We also have similar
evidence from quantum theory: the ultraviolet problems of general
relativity in harmonic gauge are well-known, highly probable they
appear in post-relativity too.  That's why we make the following
``atomic hypothesis'':

\begin{center}{\it
Post-relativistic gra\-vity is only the large scale approximation of a
different, atomic ether theory.}
\end{center}

Note that this atomic hypothesis destroys the relativistic symmetry of
post-relativity in the domain $\lambda_1,\lambda_2\approx 0$.  The
atomic theory has a completely different symmetry group compared with
the large scale approximation.  Relativistic time dilation is only a
large scale effect.  The atomic theory should be Galilean invariant,
but there will be no local Lorentz invariance.  Thus, for distances
where the atomic structure of the ether becomes observable,
relativistic symmetry becomes invalid, the hidden ``preferred frame''
becomes observable.

There are a lot of different possibilities for ``atomic models''.  All
what follows from the internal problems of continuous theory is a
breakdown of this theory for absolute distances below some
$\epsilon>0$.  But even this $\epsilon>0$ is unknown yet.

To find out if the atomic hypothesis allows to solve these internal
problems, the details of the atomic model seem to be not interesting.
For this purpose, a simple example theory should be sufficient.  Let's
define now such a theory.  Note that the purpose of this theory is not
so much to describe a nice atomic model, but to have a simple example
theory for computations.

 \begin{definition} Assume we have a Galilean invariant classical
field theory with a Lagrange density $L_c$ which does not depend on
higher than first derivatives of the fields $\phi$.

The ``related lattice theory'' is defined by the following steps of
freedom:

 \begin{itemize}

 \item a regular rectangular 3D lattice with distance $\epsilon$
between the nodes, with N nodes in each direction;

 \item for every field $\phi(x)$ of the continuous theory one step
of freedom $\phi_x$ for every node x of the lattice;

 \end{itemize}

and it's Lagrange density 

\[ L = L_{lattice} + L_c(i(\phi_x)) \]

Here $L_{lattice}$ denotes the Newtonian Lagrangian of a rigid body
for the position of the lattice as a whole, and $i(\phi_x)$
denotes the standard (tri-linear) first order finite element
embedding.
\end{definition}

The finite element embedding $i(\phi_x)$ is a function with
$(i(\phi_x))(y) = \phi_y$ for every node y, interpolated for the other
points.  It is a continuous function with discontinuous first order
derivatives.  That's why we include the condition that $L_c$ does not
depend on higher than first derivatives.

The movement of the lattice as a whole may be ignored. The only
purpose of this construction was to make the lattice theory
Galilean invariant. 

The most interesting example is post-relativistic gra\-vity.  The
related lattice theory we name post-relativistic lattice theory.  To
fit into the scheme, we have to use the Rosen Lagrangian instead of
the curvature as the relativistic Lagrangian $L_{rel}$.  This
well-known Lagrangian

\[ L_{Rosen} = g^{ik}
	(\Gamma^m_{il}\Gamma^l_{km}-\Gamma^l_{ik}\Gamma^m_{lm})
\]

does not depend on second order derivatives, but leads to the same
relativistic equations like scalar curvature R.

Thus, we have defined a simple example theory with a finite number of
steps of freedom.  It fulfills the necessary requirements of our
approach: Galilean invariance and agreement with experiment.

\subsection{Better Atomic Ether Theories}

Post-relativistic lattice theory is only one example theory with a
microscopic structure.  It fulfils the necessary requirements ---
Galilean invariance and agreement with experiment.  From mathematical
point of view, lattice theory seems to be the simplest discrete
theory.

But there is an interesting class of theories which has to be
preferred because of higher predictive power --- theories which really
justify the notion ``atomic models''.

Indeed, the steps of freedom of the ether highly remember classical
matter, and we have classical conservation laws.  These properties
have not been used in lattice theory.  Better atomic models with the
following properties:

 \begin{itemize}

 \item Galilean invariance,

 \item atoms as particles with a certain position as steps of freedom,

 \item the ether density $\rho$ and velocity $v^i$ of continuous ether
theory as particle density and average velocity of these particles.

 \end{itemize}

automatically explain these properties.  Some other ideas for atomic
theories:

 \begin{itemize}

 \item There may be a crystal structure.  The axes may be described by
a triad formalism.  This is a natural modification of the relativistic
tetrad formalism.  We already have a predefined splitting into space
and time direction, thus, we need only three vector fields in space.

 \item Visible fermions may be interstitials, their anti-particles
vacancies.  That's a variant of Dirac's original idea.

 \item Gauge fields describe different types of deformations related
with these crystal defects.  It is known that gauge formalism may be
used to describe crystal defects.

 \end{itemize}

Thus, even without experimental support, further improvement of the
atomic ether models is possible.  The ideal final result may be
something like a qualitative crystal model of the ether which explains
the observed particles and gauge fields as some types of crystal
defects.  This will be the ether-theoretical replacement of a unified
field theory.  But, of course, even in this ideal picture we have a
lot of unexplained remaining parameters, like the masses of the
different sorts of ether atoms.

\section{Canonical Quantization}

Now, let's show that the hypothesis of a different microscopic
structure really allows to solve the quantization problem.  For this
purpose, it is sufficient to show the existence of a quantum theory
with the necessary properties: Galilean invariance and agreement with
experiment.  Moreover, it is sufficient to do this for our simple
example theory --- post-relativistic lattice theory.

\begin{theorem} Assume we have a classical Galilean invariant theory
with Lagrange formulation so that the Lagrangian does not depend on
higher than first derivatives of the variables.

In this case, there exists a quantum theory with this theory as 
some classical approximation.
\end{theorem}

This is straightforward: we can apply classical canonical quantization
to our theory.  Indeed, we have a theory with a finite number of steps
of freedom, and a classical Lagrange mechanism with a Lagrange density
which does not depend on higher than first derivatives of these steps
of freedom.

This is already sufficient to derive the Hamilton formulation in the
canonical way.  Constraints we remove by small regularization terms
like $\epsilon\dot{\lambda}^2$ for a Lagrange multiplier $\lambda$.

Once we have found a Hamilton formulation, the only remaining
quantization problem is the definition of the Hamilton operator
$\hat{h}$ for the classical Hamilton function $h(p,q)$.  But the
existence can be proven for arbitrary $L^{\infty}$-functions.  If Weyl
quantization (which works for $L^2$) fails we can use anti-normal
quantization (which works for $L^\infty$).  If even this scheme fails
because $h(p,q)$ is not bounded, we regularize it using some large
enough energy $H_0$ for cutting.  This does not change Galilean
invariance and agreement with experiment. qed.

Note that the regularizations we have allowed in this general scheme
are modifications of the classical Hamilton formalism.  The
quantization itself is pure, standard, canonical quantization.

Now we can apply this scheme to post-relativistic lattice theory and
obtain the main result:

\begin{theorem} There exists a Galilean invariant quantum theory of
gra\-vity, obtained by classical canonical quantization, which is in
agreement with experiment in the relativistic domain, even for strong
gravitational fields.  \end{theorem}

Of course, there may be better variants.  Instead of a regularization
of the constraints, we will obviously prefer the generalized Hamilton
formalism introduced by Dirac.  But all we need for the purpose of
this paper is to show the existence of such a quantum theory.

\section{Discussion}

This shows that special and general relativistic effects are
compatible with Galilean invariance, the metaphysics of
Lorentz-Poincare ether theory and canonical quantization.  

Our argumentation is very simple.  The mathematical part of theory we
have presented is trivial --- de-facto a single sentence.  We apply a
methodological rule which is also a single sentence: we have to choose
the best available theory.  The conclusion is very non-trivial: we
have to choose this non-relativistic theory as the best available
theory of quantum gra\-vity.  That means, the relativistic paradigm
has to be rejected.  Instead, we have to use pre-relativistic
ether-theoretical metaphysics.

It seems, the simplicity of our argumentation is essential.  Only a
very simple argumentation which does not leave place for loopholes and
immunization has the power to destroy such a fundamental paradigm like
relativity.

We have to remark that it is not our simple theory which invalidates
relativity.  What we have done was to prove that there is no
compatibility problem between gra\-vity, Lorentz-Poincare ether theory
and canonical quantization.  Relativity has to be rejected because it
is incompatible with quantum gra\-vity.  There is enough support for
the incompatibility hypothesis: the problem of time \cite{Isham},
topological foam, the information loss problem, non-renormalizability.
It's time to draw the conclusions and to reject relativity.

Moreover, this is not the only reason to reject relativity.  The other
is an experimental result --- the violation of Bell's inequality by
Aspect's experiment \cite{Aspect}.  If we accept the EPR argumentation
\cite{EPR}, it is a clear experimental violation of Einstein
causality.  Bell's conclusion was (cited by \cite{Price94}):

{\it ``the cheapest resolution is something like going back to
relativity as it was before Einstein, when people like Lorentz and
Poincare thought that there was an aether --- a preferred frame of
reference --- but that our measuring instruments were distorted by
motion in such a way that we could no detect motion through the
aether. Now, in that way you can imagine that there is a preferred
frame of reference, and in this preferred frame of reference things go
faster than light.'' }

At this time, there was a strong counter-argument: to go back to the
Lorentz ether means to throw away general relativity --- one of the
most successful theories of our century. Indeed, in general relativity
we have solutions with non-trivial topology, but this is incompatible
with Lorentz ether theory.  In other words, Lorentz ether theory was
much more wrong than general relativity, because it was unable to
describe gravity.

Now, by the way, this counter-argument is removed.  We have an ether
theory of gra\-vity which is compatible with relativistic experiment.
Thus, Bell's conclusion is now much stronger, we have de-facto no
costs if we ``go back to relativity as it was before Einstein''.

This independent argument against relativity remains valid even if a
relativistic quantum theory of gra\-vity will be found in future
following one of the many research directions (\cite{Anandan},
\cite{Ashtekar}, \cite{Baez}, \cite{DeWitt}, \cite{Kuchar},
\cite{Isham}, \cite{Hartle}, \cite{Hawking79}, \cite{Thiemann}).  This
theory should not only solve all of the well-known problems, but it
should present obvious advantages to overweight their problem with the
violation of Bell's inequality.  From this point of view, the
relativistic paradigm is as dead as possible for a scientific
paradigm.

\subsection{Conclusion}

According to our argumentation, the relativistic paradigm has to be
rejected.  We have two independent reasons for this rejection: The
non-existence of a relativistic quantum theory of gra\-vity, and the
violation of Bell's inequality.

We have to replace relativity by a generalization of Lorentz-Poincare
ether theory to gra\-vity which does not have these faults.  It
follows that:

 \begin{itemize}

 \item Our universe is Galilean invariant.  There exists an absolute
space and absolute time.  The space is filled with an ether ---
another word for the gravitational field.

 \item Our clocks and rulers are distorted by interaction with the
ether.  

 \item Our universe is flat and not expanding.  Only our rulers become
smaller.  There is no big bang in the past in absolute time.

 \item Black holes do not exist, instead we have ``frozen stars''.  In
absolute time, the collapse stops immediately before horizon
formation.  The part ``behind the horizon'' is not physical.

 \end{itemize}

The assumption that our current continuous ether theory remains valid
at arbitrary small distances has to be rejected.  The ether probably
has a yet unknown atomic structure.

For this atomic theory we can apply canonical quantization.  This has
been shown for a simple discrete example theory, and there is no
reason to doubt that it works for other atomic ether theories too.


\begin{thebibliography}{99}

\bibitem{Anandan}
J.~Anandan, Gravitational phase operator and cosmic string,
Phys. Rev. D53, nr. 2, p.779-786, 1996

\bibitem{Ashtekar}
A. Ashtekar, Recent Mathematical Developments in Quantum General
Relativity, gr-qc/9411055, 1994

\bibitem{Aspect}
A. Aspect, J. Dalibard, G. Roger, Experimental test of Bell's
inequalities using time-varying analysers, Phys. Rev.Lett. 49,
1801-1807, 1982

\bibitem{Baez} J.C.~Baez, Spin Networks in Nonperturbative Quantum
Gra\-vity, to appear in the proceedings of the AMS Short Course on
Knots and Physics, San Francisco, Jan. 2-3, 1995, ed. Louis Kauffman

\bibitem{Bell}
J.S.~Bell, Speakable and unspeakable in quantum mechanics, Cambridge
University Press, Cambridge, 1987

\bibitem{DeWitt}
B.S. DeWitt, Quantum Gra\-vity: New Synthesis, in S. Hawking (ed.)
General Relativity, Cambridge University Press 1979

\bibitem{EPR}
A.~Einstein, B.~Podolsky, N.~Rosen, Can quantum-mechanical description
of physical reality be considered complete?, Phys. Rev. 47, 777-780,
1935

\bibitem{Fock}
V.~Fock, Theorie von Raum, Zeit und Gravitation, Akademie-Verlag
Berlin, 1960

\bibitem{Hartle} J.B. Hartle, Unitarity and Causality in Generalized
Quantum Mechanics for Non-Chronal Spacetimes, gr-qc/9309012,
Phys. Rev. D49 6543-6555, 1994

\bibitem{Hawking79}
S.W.~Hawking, in: General relativity Cambridge University Press,
1979

\bibitem{Isham} C.~Isham, Canonical Quantum Gra\-vity and the Problem
of Time, Presented at 19th International Colloquium on Group
Theoretical Methods in Physics, Salamanca, Spain, gr-qc/9210011, 1992

\bibitem{Kuchar} K.V.~Kuchar, C.G.~Torre, Harmonic gauge in canonical
gra\-vity, Physical Review D, vol. 44, nr. 10, pp. 3116-3123, 1991

\bibitem{Lanczos}
K.~Lanczos, Ein vereinfachendes Koordinatensystem fuer die
Einsteinschen Gravitationsgleichungen Phys. Z., 23, 537, 1922

\bibitem{Popper}
K.R. Popper, Conjectures and Refutations, Routledge \& Kegan Paul,
London, 1963

\bibitem{Price94}
H. Price, A neglected route to realism about quantum mechanics, Mind
103, 303-336, gr-qc/9406028

\bibitem{Schmelzer92}
I.~Schmelzer, Quantization and measurability in gauge theory and
gra\-vity, WIAS preprint nr. 18, ISSN 0942-4695, Berlin, 1992

\bibitem{Schmelzer96a} I.~Schmelzer, Postrelativistic gra\-vity - a
hidden variable theory for general relativity, gr-qc/9605013, 1996

\bibitem{Schmelzer96b}
I.~Schmelzer, Postrelativity - a paradigm for quantization with
preferred Newtonian frame, WIAS preprint n. 286, Berlin, also
gr-qc/9610047, 1996

\bibitem{Sokal} A.~Sokal, Transgressing the Boundaries: Towards a
Transformative Hermeneutics of Quantum Gra\-vity, Social Text 46/47
pp. 217-252, 1996

\bibitem{Thiemann}
T.~Thiemann, gr-qc/9705017 -- gr-qc/9705021, 1997

 \end{thebibliography}
\end{document}